\documentclass[11pt, reqno]{article}
\usepackage{graphicx,amsmath,amssymb,epsfig}
\usepackage{amsfonts, lscape, harvard}
\usepackage{amsthm}
\usepackage{color,url}
\usepackage{bm, multirow}
\usepackage{hyperref}
\hypersetup{
    colorlinks=true,
    linkcolor=blue,
    filecolor=magenta,
    urlcolor=cyan,
}

\theoremstyle{plain}

\numberwithin{equation}{section}
\input{tcilatex}

\def\uniset{{\rm 1\kern-.40em 1}}

\newcommand{\abs}[1]{\lvert#1\rvert}

\begin{document}
\title{{\large \bf The Econometrics and Some Properties of Separable Matching Models}}
\author{{\bf Alfred Galichon}\thanks{Department of Economics, FAS, and Department of
Mathematics, Courant Institute, New York University. Email:  ag133@nyu.edu.}
 \and 
{\bf Bernard Salani\'{e}}\thanks{
Department of Economics, Columbia University. Email: bs2237@columbia.edu.}}
\date{December 28, 2016}

\maketitle

\begin{abstract}
	We present a class of one-to-one matching models with perfectly  transferable utility. We discuss  identification and inference in these separable models, and we show how their comparative statics are   readily analyzed.  
\end{abstract}


{\footnotesize \textbf{JEL Classification}: D3, J21, J23 and J31.\vskip50pt }

\section{\setcounter{page}{1}\setcounter{equation}{0}Introduction}

Eugene Choo and Aloysius Siow's (2006)  contribution has renewed 
interest in empirical applications of matching with perfectly transferable utility (TU). 
 Unobserved
heterogeneity in joint surplus is a paramount consideration in the specification of these models. Choo and Siow chose a  separable multilogit model, which leads to  highly tractable formul\ae. But  
unobserved heterogeneity
could originate from variation in tastes, from division of labor within the partners, and other sources. It is therefore important to allow for flexibility in the stochastic specification of the joint surplus.
 Alfred Galichon and Bernard Salani\'{e} (2016) have explored a general class
of  models of bilateral matching which sets very few constraints on the  distributions of  unobserved heterogeneity beyond {\em separability\/} of the joint surplus.  These separable models have a nicely convex and (usually) smooth structure  that generates 
very useful econometric and analytic properties.

We start by summarizing our main results concerning identification and inference in separable models of one-to-one matching under TU\footnote{Galichon--Salani\'e (2016) has detailed arguments, along with somehat weaker assumptions than we use here.}. We then show how 
in models with separable heterogeneity and full support, we  can use the implicit function theorem and matrix algebra to get explict formul\ae\ for any small change in the primitives of the model:  arrival or departure of a mass of individuals of a given type, or changes in joint surplus.
We illustrate the usefulness of our formul\ae\ on a simple example.

\section{Separable models with full support} 
\label{sec:separable_models_with_full_support}
In this paper we will call ``men'' and ``women'' the agents on both sides of the market, as is traditional; but our results apply more generally than in  this implicit heterosexual marriage market \`a la Becker. 
We  assume that agents on both sides of the market belong to continuous sets $\mathcal{I}$ and $\mathcal{J}$, which  are partitioned into finite sets of types. 
A man $i\in \mathcal{I}$ has a type $x_i\in \mathcal{X}$ and a woman $j\in\mathcal{J}$ has a type $y_j\in\mathcal{Y}$, where $\mathcal{X}$ and $\mathcal{Y}$ are finite. 
The mass of men of type $x$ (resp.\ women of type $y$) is $n_x$ (resp.\ $m_y$).
The distinction between types and identities is data-driven: while participants on the market are assumed to  operate under perfect information, the analyst only observes the types $x$ and $y$. 
  We also assume that joint surplus  is {\em separable}:
\[
\tilde{\Phi}_{ij}=\Phi_{x_i y_j}+\varepsilon_{iy_j}+\eta_{jx_i}.
\]
Separability excludes interactions between unobserved characteristics  of $i$ and $j$ conditional on observed types $(x,y)$. As an example, let types describe education, as in Pierre--Andr\'e Chiappori, Salani\'e, and Yoram Weiss (2016). Then separability does allow for unlimited unobserved heterogeneity in  the way more-educated men value the education of their partners for instance; it rules out considerations like matching on physical characteristics, which certainly exists but may not be that relevant for the study of some economic questions at least. 

We know from Chiappori-Salani\'e-Weiss (2016) and Galichon--Salani\'e (2016) that 
 if $i$ and $j$ match  then the man receives utility $%
U_{x_iy_j}+\varepsilon _{iy_j}$ and the woman receives utility $V_{x_iy_j}+\eta _{jx}$,
where the terms  $\bm{U}=(U_{xy})$ and $\bm{V}=(V_{xy})$ are endogenously determined
at equilibrium so that $U_{xy}+V_{xy}=\Phi _{xy}$.  A single man $i$  receives utility $\varepsilon _{i0}$, while a single woman $j$  receives $\eta _{j0}$.   The interpretation of this result is simple:  the $\varepsilon_{iy}$ of man $i$ has the same value for ll women of type $y$, and since there is a continuum of them  they will compete for it until the ``price'' of man $i$ fully incorporates it. 

We now denote $\mathcal{X}_0=\mathcal{X} \cup \{0\}$ and  $\mathcal{Y}_0=\mathcal{Y} \cup \{0\}$.
 We shall assume that 
the random vector $\bm{\varepsilon}_x =(\varepsilon_{iy})\in \mathbb{R}^{\mathcal{Y}%
_{0}}$ is distributed as $\mathbf{P}_{x}$ identically and independently across the population of men $i$ of type $x$;  and we introduce $\mathbf{Q}_{y}$ in the same way for women. 
In this note we will also impose {\em full support}:  for each $%
x\in \mathcal{X}$, $\mathbf{P}_{x}$ has a nonvanishing density on $\mathbb{R}%
^{\mathcal{Y}_{0}}$,  and for each $y\in \mathcal{Y}$, $\mathbf{Q}_{y}$ has
a nonvanishing density on $\mathbb{R}^{\mathcal{X}_{0}}$. 

\section{Equilibrium and welfare} 
\label{sec:equilibrium_and_welfare}

 When $\mathbf{P}%
_{x}$ and $\mathbf{Q}_{y}$ are Gumbel distributions for all $x$ and $y$, the
model boils down to the model of Choo and Siow (2006). More generally,
Galichon and Salani\'{e} (2016) introduce the convex functions
\begin{equation*}
G_{x}\left(\bm{U}\right) =\mathbb{E}_{\mathbf{P}_{x}}\left[ \max_{y\in \mathcal{Y%
}}\left\{ U_{xy}+\varepsilon _{y},\varepsilon _{0}\right\} \right] \text{
and }H_{y}\left(\bm{V}\right) =\mathbb{E}_{\mathbf{Q}_{y}}\left[ \max_{x\in 
\mathcal{X}}\left\{ V_{xy}+\eta _{x},\eta _{0}\right\} \right],
\end{equation*}%
and 
\begin{equation*}
G\left(\bm{U}\right) =\sum_{x\in \mathcal{X}}n_{x}G_{x}\left(\bm{U}\right) \text{
and }H\left(\bm{V}\right) =\sum_{y\in \mathcal{Y}}m_{y}H_{y}\left(\bm{V}\right).
\end{equation*}%
Galichon and Salani\'{e} (2016, Theorem~2) show that $\bm{U}$ minimizes  the expression 
$G\left(\bm{U}\right) +H\left(\bm{\Phi} -\bm{U}\right)$. Under separability and full support, the functions $G$ and $H$ are 
strictly convex and twice differentiable, and 
the first-order conditions characterize the unique equilibrium:
\begin{equation}
\nabla G\left(\bm{U}\right) =\nabla H\left(\bm{\Phi} -\bm{U}\right).   \label{equil}
\end{equation}
These conditions are easily interpreted. 
By the Daly-Zachary-Williams theorem,  the
mass of men of type $x$ wishing to match with women of type $y\in \mathcal{Y}
$ given a vector $\bm{U}$  is $\bm{U}$ is  $\mu _{xy}=\partial G\left(\bm{U}\right)/\partial U_{xy}$.
Similarly, the number of women of type $y$ wishing to match with men of type 
$x\in \mathcal{X}$ is $\mu_{xy} =\partial H\left(\bm{V}\right)/\partial V_{xy}$. In
equilibrium, the two quantities $\nabla G\left(\bm{U}\right)$ and $\nabla
H\left(\bm{V}\right)$ must coincide; and  since $\bm{U}+\bm{V}=\bm{\Phi}$,
$\bm{U}$ is determined in equilibrium by~\eqref{equil}. Also note that the expected utility of the average man of type $x$ 
 is $u_x=G_x(\bm{U})$ in equilibrium.
Galichon--Salani\'e (2016, section 5) details several approaches to computing the equilibrium efficiently. The convexity and smoothness of the problem make it very tractable numerically.


\section{Identification and Inference} 
\label{sec:identification_and_inference}
Convex duality is the key to the approach in Galichon and Salani\'e (2016). Remember that given any function $f(a)$, its Legendre--Fenchel transform is the function $f^\ast$ such that
\[
f^\ast(b)=\sup_a \left(a\cdot b-f(b)\right).
\]
The function $f^\ast$ may be badly-behaved: it may take infinite values, for instance. But  since $f^\ast$ is the supremum of linear functions of $b$,  it is convex. And if $f$ is convex, it is the Legendre--Fenchel transform of $f^\ast$; and if $f$ and $f^\ast$ are strictly convex, then  
\[
b=\nabla f(a) \; \mbox{ iff } \; a=\nabla f^\ast(b).
\]
Let us first  apply this ``convex inversion formula''  to the strictly convex function $f=G$: 
\[
\bm{\mu}=\nabla G(\bm{U})  \; \mbox{ iff } \; \bm{U}=\nabla G^\ast(\bm{\mu}).
\] 
Given a full specification for  the distributions $\mathbf{P}_x$, the function $G$ can be computed, and its Legendre--Fenchel transform too. Feeding the observed matching patterns into $\bm{U}=\nabla G^\ast(\bm{\mu})$ directly identifies $\bm{U}$.  Proceeding in the same way with $f(\bm{U})=H(\bm{\Phi}-\bm{U})$ identifies $\bm{\Phi}-\bm{U}=\nabla H^\ast(\bm{\mu})$; and adding up,
\[
\bm{\Phi}=\nabla G^\ast(\bm{\mu})+\nabla H^\ast(\bm{\mu}),
\]
which identifies the joint surplus $\bm{\Phi}$ from the (assumed) knowledge of the distributions $\mathbf{P}_x$ and $\mathbf{Q}_y.$

This is ``conditional unrestricted identification'': the joint surplus is identified without any prior restriction if  the 
analyst somehow knows the distribution of unobserved heterogeneity. If for instance these distributions are only assumed to be known up to scale, then in order to achieve point identification of the joint surplus the analyst will need to impose restrictions on it. There is an unavoidable trade off here, which can be alleviated by pooling data from several markets and assuming some common features across markets\footnote{Chiappori, Salani\'e and Weiss (2016) gives an  example, with an heteroskedastic version of the Choo and Siow model.}.

Once identification is achieved, inference is straightforward. It can be based directly on the equations above, or proceed via maximum likelihood, or by matching moments of some basis functions. The latter method is based on a linear expansion 
\[
\Phi_{xy}(\bm{\lambda})=\lambda\cdot \bm{\phi}_{xy}
\]
where $\bm{\phi}$ is a vector of basis functions\footnote{For instance, a simple ``assortative matching basis function'' would be $\uniset(x=y)$.}. Galichon and Salani\'e (2016) show that finding the parameter vector $\bm{\lambda}$ that matches the observed {\em comoments\/} $\hat{\bm{C}}=\hat{E} \bm{\phi}$ gives a consistent estimator.


\section{Comparative Statics}
The separable structure of the  problem naturally generates a number of comparative statics results that extend those obtained by Colin Decker et al.~(2012) and  Bryan Graham~(2013) for the Choo and Siow model. Our assumptions on the unobserved
heterogeneity yield enough smoothness and convexity that simple
formul\ae\ can be obtained.

Take a well-known result:  in two-sided matching models, the arrival of newcomers on one side of the
market hurts all participants on the same side of the market, and benefits
all participants on the opposite side of the market. This was proved by Alexander Kelso and Vincent Crawford (1982, Theorem~5) for a many-to-one matching model under  a gross substitutes assumption; by David Gale and Marilda Sotomayor (1985, Theorem~2) for the NTU marriage model; and by Gabrielle Demange and  Gale (1985, Corollary~3) for a general class of one-to-one models with transfers. But all of these proofs are purely qualitative. With separable models, it is easy to make these results quantitative, and more generally to analyze the effects of small changes in the primitives.

The functions $G$ and $H$ are not only twice differentiable and strictly convex: they are also
 submodular. The 
economic interpretation is straightforward. Given differentiability, the submodularity of $G$
requires that $\partial G^{2}/\partial
U_{xy}\partial U_{x^\prime y^{\prime }}\leq 0$ for  all $(x^\prime,y^{\prime})\neq (x,y)$. But since $\mu _{xy}=\partial G_x\left(
\bm{U}\right)/\partial U_{xy}$, this simply says  that $\partial \mu _{xy}/\partial
U_{xy^{\prime}}\leq 0$: if alternative $y^{\prime}$
becomes more attractive,  alternative $y$  will be
less demanded at equilibrium. This is, of course, a gross substitutes property.

\bigskip 

To state our results, we need some more notation:
\begin{itemize}
	\item we define matching ratios by $\mu_{xy}=\mu^M_{y\vert x}n_x=\mu^W_{x\vert y}m_y$;  note that 
	$\sum_y \mu^M_{y\vert x}=1-\mu^M_{0\vert x}$ and $\sum_x \mu^W_{x\vert y}=1-\mu^W_{0\vert y}$.
	\item  we denote $\bm{T}=\left(D^{2}G\left(\bm{U}\right) +D^{2}H\left(\bm{\Phi}-\bm{U}\right)\right)^{-1}$ the inverse of the sum of  the Hessians of $G$ and $H$ at the equilibrium $\bm{U}$ (the sum is invertible since $G$ and $H$ are strictly convex.) 
	\item We   use specific notation for some of its blocks; for instance, we denote
	$\bm{T}_{x\cdot,\cdot y}$ the matrix $\bm{A}$ with elements $A_{tz}=T_{xt,zy}$. 
\end{itemize}

\subsection{General results for separable models} 
\label{sub:general_results_for_separable_models}
	The primitives of the model are $\bm{\theta}=(\bm{n},\bm{m},\bm{\Phi})$. 
The equilibrium $\bm{U}$ is determined by $\nabla G\left(\bm{U}\right) =\nabla
H\left(\bm{\Phi} -\bm{U}\right)$.  Taking differentials, for all $x$ and $y$ we have
\begin{equation}
	\label{eq:dtheta}
\left\{D^{2}G\left(\bm{U}\right) +D^{2}H\left(\bm{\Phi}-\bm{U}\right) \right\} d\bm{U}=
\frac{\partial^2 H\left(\bm{\Phi}-\bm{U}\right)}{\partial \bm{V}\partial \bm{m}}d\bm{m}
-
\frac{\partial^2 G\left(\bm{U}\right)}{\partial \bm{U}\partial \bm{n}}d\bm{n}
+D^{2}H\left(\bm{\Phi}-\bm{U}\right)d\bm{\Phi}.
\end{equation}%
Given strict convexity, the Hessians are negative definite, 
and the matrix $D^{2}G\left(\bm{U}\right) +D^{2}H\left(\bm{\Phi}-\bm{U}\right)$ is invertible.
Therefore we can write
\begin{equation}
d\bm{U}=\bm{T} \bm{R}d\bm{\theta},
\end{equation}
where $\bm{R}d\bm{\theta}$ denotes the right-hand side of \eqref{eq:dtheta}.
Now since 
both $G$ and $H$ are submodular and strictly convex, $D^{2}G$ and $D^{2}H$
are Stieltjes matrices\footnote{That is, they are positive definite with non-positive off-diagonal terms.},  and so is  their
sum. By a classical result on Stieltjes matrices
(see e.g. Golub and Van Loan 2013, lemma 11.5.1), all entries of $\bm{T}$
are nonnegative; and any change in $\bm{\theta}$ such that $\bm{R}d\bm{\theta}$ is a non-negative vector can only increase the equilibrium  $U_{xy}$.  Moreover, 
the average welfare of men of type $x\in \mathcal{X}$ is given by $%
u_x=G_{x}\left(\bm{U}\right)$, and
\[
du_x=\sum_{y\in \mathcal{Y}}\frac{\partial G_{x}}{\partial U_{xy}}dU_{xy}=\sum_{y\in \mathcal{Y%
}}\mu^M_{y\vert x}dU_{xy};
\]
so that any such change $\bm{R}d\bm{\theta}\geq 0$ can only increase the average expected utilities of men of any type.

Applying this to small changes in population sizes $\bm{n}$ and $\bm{m}$  yields  very simple  formul\ae\footnote{The online appendix has the detail of these calculations.}:
\begin{align}
 \frac{\partial u_x}{\partial n_{x^{\prime }}} & =\bm{\mu}^M_{\cdot\vert x^\prime} \bm{T}_{x\cdot,x^\prime\cdot} \bm{\mu}^M_{\cdot\vert x^\prime}\leq 0 \notag \\
\frac{\partial  u_x}{\partial m_{y^{\prime }}} &=\bm{\mu}^M_{\cdot\vert x^\prime} \bm{T}_{x\cdot,\cdot y^\prime} \bm{\mu}^W_{\cdot\vert y^{\prime}} \geq 0.
\end{align}
The signs of the entries is a direct consequence of the non-negativity of all elements of $\bm{T}$; it was already known, but now we can easily compute the value of these local effects. In addition, it is easy to prove that
\[
\frac{\partial u_x}{\partial \Phi_{x^\prime y^\prime}}  =\mu_{xy^\prime}\bm{T}_{xy^\prime,\cdot y^\prime}
\frac{\partial^2 H_{y^\prime}}{\partial V_{x^\prime y^\prime}\partial V_{\cdot y^\prime}}.
\]
Since  $H_{y^\prime}$ is strictly convex and is submodular, the vector of second derivatives in this expression has one positive term, while  all others are non-positive. Given the non-negativity of all elements of $\bm{T}$,  an  increase in any element $\Phi_{x^\prime y^\prime}$ of the joint surplus  should reduce (resp.\ increase) the expected utility of men whom women   of type $y^\prime$ see as  good (resp.\ bad) substitutes of type $x^\prime$. These effects are larger for the  men who are more likely to marry women of  type $y^\prime$.

More generally, for any small change in the primitives of the model, we recover $du_x=\sum_y \mu^M_{y\vert x} dU_{xy}$ from the solution of the system
\begin{align}
\label{eq:full}	
& n_x\sum_t \frac{\partial^2 G_x}{\partial U_{xy}\partial U_{xt}} dU_{xt}+
	m_y\sum_z \frac{\partial^2 H_y}{\partial V_{xy}\partial V_{zy}}dU_{zy} \notag \\
& \; \; =	 \mu_{xy}d\log\frac{m_y}{n_x}+m_y\sum_z \frac{\partial^2 H_y}{\partial V_{xy}\partial V_{zy}}d\Phi_{zy}.
\end{align}
While $G_x$ and $H_y$ are functions of $\bm{U}$,  using the Legendre-Fenchel transform we have $\bm{\mu}=\nabla G^\ast(\bm{\mu})+H^\ast(\bm{\mu})$.
Hence all of the elements of~\eqref{eq:full} can be computed from  the observed data, given a structure $(\bm{\Phi},\bm{n},\bm{m}).$


\subsection{A one-type model} 
\label{sub:a_one_type_model}
For a drastically simple illustration, suppose that there is only one type of men and one type of women: $\abs{X}=\abs{Y}=1.$ We simplify the notation by dropping the ``1'' subscripts, so that $\Phi$ denotes $\Phi_{11}$ for instance. Equilibrium in this model consists in a number of marriages $\mu$, and associated expected utilities $u$ and $v$.

Now $G(U)=nE_{\mathbf{P}} \max(U+\varepsilon,\varepsilon_0)$. Let us  denote $(F_P, f_P)$ the cdf and pdf of $(\varepsilon_0-\varepsilon)$ under $\mathbf{P}$; and define $k_P(t)=f_P(F_P^{-1}(t))$. Then $G^\prime(U)=nF_P(U)$ and $G^{\prime\prime}(U)=nf_P(U).$ Using similar notation for $\mathbf{Q}$, the equilibrium $U$ and the number of marriages $\mu$ are given by $\mu=nF_P(U)=mF_Q(\Phi-U)$.  Identification is straightforward: given $\mathbf{P}$ and $\mathbf{Q}$,  solving these equations for $\Phi$ gives 
\[
\Phi=F_P^{-1}\left(\frac{\mu}{n}\right)+F_Q^{-1}\left(\frac{\mu}{m}\right).
\]
Moving to comparative statics, \eqref{eq:full} becomes
\[
dU=T\left(\mu d\log\frac{m}{n}+m k_Q\left(\frac{\mu}{m}\right)d\Phi\right)
\]
with $T=1/S$ and $S=n k_P(\mu/n)+m k_Q(\mu/m)$. Since $du=(\mu/n)dU$, the change in the expected utilities of the average man follows directly, and so does the change in the number of marriages since $d\mu=F_P(U) dn+nf_P(U)dU:$
\begin{align}
	\label{eq:du1}
	du &= T\frac{\mu}{n}\left(\mu d\log\frac{m}{n}+m k_Qd\Phi\right) \\
		\label{eq:dmu1}
		d\mu &= T\left(\mu \left(m k_Q d\log n+n k_P  d\log m\right)+nm k_P k_Q d\Phi\right).
\end{align}
Take a small change $(dn,dm)$ in the sizes of the populations of men and of women.  The resulting log-change $d\log\mu$ in the number of marriages will be a weighted average of the log-changes  in $n$ and in $m$. More interestingly,

\medskip

{\em the changes in expected utilities of men and women directly reflect the change in the sex ratio $n/m$; and so do the changes in the percentage of singles in each gender.}

\medskip

If only the joint surplus of each marriage changes, by $d\Phi$, then

\medskip

{\em the  number of marriages $\mu$ changes by a fraction $0<s<1$ of $d\Phi/2$.}

\medskip

Assume moreover $(\varepsilon_0-\varepsilon)$ and $(\eta_0-\eta)$ have the same distribution, with cdf $F$ and pdf $f$; and let it be symmetric around 0 and log-concave.  Then $0<U<\Phi/2$ and $v>u>0$  if  the sex ratio is unfavorable to men, $n<m$. Log-convavity gives us $k_P(\mu/n)<k_Q(\mu/m)$, so that $nk_P< mk_Q$ and $Tnk_P<1/2<Tmk_Q.$  Therefore

\medskip

{\em the number of marriages is more elastic to the  size of the smaller population.}

\medskip


\newpage

\section*{References} 
\ \\

Chiappori, P.-A.,  Salani\'e, B. and  Y. Weiss (2016): ``Partner Choice, Investment in Children, and the Marital College Premium'',
mimeo.

Choo, E., and A. Siow (2006): \textquotedblleft Who Marries
Whom and Why,\textquotedblright\ \emph{Journal of Political Economy},
114, 175--201.

 Decker, C., E. Lieb, R. McCann, and B.
Stephens (2012): \textquotedblleft Unique Equilibria and Substitution
Effects in a Stochastic Model of the Marriage Market,\textquotedblright\
\emph{Journal of Economic Theory}, 148, 778--792.

  Demange, G. and D. Gale (1985):
 ``The Strategy Structure of Two-Sided Matching Markets '', \emph{Econometrica}, 53,  873--888.

Gale, D. and M. Sotomayor (1985): ``Some Remarks on the Two-sided Matching Model'', {\em Discrete Applied Mathematics}, 11, 223--232.

 Galichon, A., and B. Salani\'{e} (2016): \textquotedblleft
Cupid's Invisible Hand: Social Surplus and Identification in Matching
Models,\textquotedblright\ working paper.

 Golub, G. and Van Loan, C. (2013). \emph{Matrix
computations}. 4th edition. Johns Hopkins.

Graham, B. (2013): ``Comparative static and computational methods for an
empirical one-to-one transferable utility matching model''. \emph{Structural
Econometric Models} 31, 153--181.

 Kelso, A. and V. Crawford (1982): ``Job Matching, Coalition Formation, and Gross Substitutes'',  \emph{Econometrica},  50, 1483--1504.


\end{document}